\documentclass[preprint,preprintnumbers]{revtex4} 

\newcommand{\be}{\begin{equation}}
\newcommand{\ee}{\end{equation}}

\newcommand{\ba}{\begin{eqnarray}}
\newcommand{\ea}{\end{eqnarray}}

\newcommand{\eV}{{\rm eV} }
\newcommand{\GeV}{{\rm GeV} }

\newcommand{\keV}{{\rm keV} }
\newcommand{\Lc}{{\cal L }}
\newcommand{\fr}{\frac{\alpha_s}{8\pi}}
\newcommand{\tb}{\overline \theta}
\newcommand{\ggb}{G \cdot \widetilde G}
\newcommand{\cp}{{CP}}
\newcommand{\agg}{{a\gamma\gamma}}

\begin{document}

\bibliographystyle{apsrev}

\preprint{UAB-FT-530}
\title{Axions and Axion-like Particles}

\author{Eduard Mass{\'o}\footnote{masso@ifae.es}}

\affiliation{Grup de F{\'\i}sica Te{\`o}rica and Institut 
de F{\'\i}sica d'Altes
Energies\\Universitat Aut{\`o}noma de Barcelona\\ 
08193 Bellaterra, Barcelona, Spain}

\begin{abstract}
I review the theoretical motivation for the axion and present
an update of the experimental status of axion searches. I finally
comment on some aspects of the physics of axion-like particles.
\vspace{1pc}
\end{abstract}

\maketitle

\section{Introduction: The strong CP problem}

Consider the QCD Lagrangian 
\ba
\Lc_{QCD} &=& - \frac{1}{4}\, G^a_{\mu\nu} G^{a\mu\nu} 
+ \bar q (i \gamma^\mu \partial_\mu - M) q \nonumber \\
          &+&  \theta \, \fr\ G^a_{\mu\nu} \widetilde G^{a\mu\nu}\ + ...
\ea
The last term, the so-called $\theta$-term,
contains the dual of the gluon field 
\be
\widetilde G^a_{\mu\nu}= (1/2) \epsilon_{\mu\nu\rho\sigma}
G^{a\rho\sigma}
\ee
The $\theta$-term is Lorentz invariant
and gauge invariant so that it should be present in ${\cal L }_{QCD}$,
but since it is proportional to $\vec E_c \cdot \vec B_c$ (color fields)
is CP-violating.  

The quark fields,
\be
q = \pmatrix{u \cr d} 
\ee
and the quark mass matrix
\be
M= \pmatrix{m_u & 0  \cr 0 & m_d}
\ee 
play an important role in the discussion of the $\theta$-term.
Let us
perform a  $U(1)_A$ rotation on a quark field, say the up quark,
\be
u \longrightarrow e^{i\alpha\gamma_5}\, u
\label{chiralrotation}
\ee
The rotation is indeed chiral since left-handed and right-handed fields
transform in a different way: $u_L \ \rightarrow\ e^{-i\alpha}u_L$, 
$u_R \rightarrow e^{i\alpha}u_R$. Under such $U(1)_A$ rotation,
the up quark mass term is not  invariant
\be
- m_u\, \bar u_L u_R +\, {\rm h.c.}
  \ \rightarrow\ - m_u\, 
e^{2 i\alpha}\, \bar u_L u_R +\, {\rm h.c.}
\ee
The conclusion is that if $m_u \neq 0$, $U(1)_A$ in
(\ref{chiralrotation}) is not a symmetry.

Naively, we could think that if $m_u=0$ we would recover this axial
 symmetry. But this is not true. 
The Noether current associated to the transformation
(\ref{chiralrotation}), $j_5^\mu$,
\be
\delta \Lc \sim \partial^\mu (\bar u \gamma_\mu \gamma_5 u) = 
\partial^\mu j_{5\mu}
\ee
is anomalous  
\be
\partial^\mu j_{5\mu} = \frac{\alpha_s}{4\pi}\, \ggb \neq 0
\ee
The $U(1)_A$ symmetry is broken by quantum effects
(Adler-Bell-Jackiw anomaly). Even in a world 
with massless quarks, a $U(1)_A$ rotation has an effect in the
$\theta$-term of the QCD Lagrangian
\be
\Lc_\cp (\theta) \equiv  \theta \fr \ggb \rightarrow 
 \Lc_\cp (\theta - 2\alpha)
\label{shiftchiral}
\ee
When indeed we have at least one massless quark,
we can set $\alpha=\theta/2$ and
the $\theta$-term would be rotated away.
But it seems that $m_u\neq 0$ in the real world
so $\Lc_\cp$ is present in the theory.

There might be still another possibility for not having $\Lc_\cp$: 
since
$\theta$ is a free parameter perhaps we could put $\theta=0$ and
forget about the $\theta$-term. However, one then has to face 
at least two problems: 1) the presence of the $\theta$-term 
is necessary to solve the
$U(1)_A$ problem (Why $\eta'$ is not a NG boson?), 
and 2) there are in fact additional contributions to $\Lc_\cp$
from the electroweak sector.
Masses generated by the electroweak spontaneous symmetry breaking 
are complex in general,
\be
\Lc_{\rm mass}= - |m_u| e^{i\varphi} \bar u_L u_R +\, {\rm h.c.} + \ldots
\ee
(the dots stand for the other quark mass terms).
To have $\Lc$ with real masses, we can use chiral rotations of the type
\be
u_L \rightarrow e^{i\varphi/2}u_L \hspace{0.75cm}
u_R \rightarrow e^{-i\varphi/2}u_R
\ee
so that
\ba
\Lc_{\rm mass} =  &-& |m_u| \bar u_L u_R +\, {\rm h.c.}+ \ldots + 
\nonumber \\
               &+&   \varphi\, \fr\ \ggb
\ea
Here, we have used (\ref{shiftchiral})
So, apart from the pure QCD $\theta$ parameter, that we may 
call $\theta_{QCD}$, 
there are additional contributions to the $\theta$-term, implying that
physics depends on
\be
  \tb = \theta_{QCD} + {\rm Arg\ Det}\ M  
\ee
The problem 2) is that even if we set $\theta_{QCD}=0$, in general 
we end up with  $\tb \neq 0$.

The most important observational consequence of $\Lc_{CP}$
is that it originates a neutron   electric dipole moment,  
\be
d_n  \sim \frac{e}{m_n}\   \tb   \ 
\frac{m_u m_d}{m_u + m_d}\,
\frac{1}{\Lambda_{QCD}}
\ee
Experimentally we have a tight bound for this observable,
\be
d_n < 0.63 \times 10^{-25}\ e\, {\rm cm}
\ee
so that we have the stringent bound
\be
  \tb < 10^{-9}  
\ee
This is the so-called strong CP-problem: Why is  $\tb$  so small?
We would have expected   $\theta_{QCD}$ and Arg Det $M$   
not far from  O(1), and we have no reason to expect
such fine-tuned cancellation between  the two terms 
$\theta_{QCD}$ and Arg Det $M$, since they have totally unrelated 
origins.

\section{The Axion and its properties}

The Peccei-Quinn (PQ) solution \cite{PecceiEM} 
to the strong CP problem
introduces a new global chiral symmetry   $U(1)_{PQ}$ 
and uses the freedom to rotate $\tb$ away. 
The spontaneous symmetry breaking of $U(1)_{PQ}$ at energy $\sim f_a$   
generates a NG boson:  the axion, $a \sim   f_a \tb$
 \cite{WeinbergEM}. We should keep in mind however that
the PQ solution to the strong CP-problem is not the unique 
solution (see \cite{ChengEM} for a review).

The axion, as all NG bosons, couples derivatively to matter 
\be
\Lc_{a\Psi\Psi} =
 \sum_i c_i\frac{1}{2   f_a } 
(\bar\Psi_i \gamma^\mu \gamma_5 \Psi_i)
(\partial_\mu a)
\ee
Here $i=e,p,n,$ etc, are the matter fields and 
$c_i=O(1)$ are model dependent parameters.
The axion is special since it has to reproduce the anomaly
and there is a non-derivative term that couples the axion
to two gluons,
\be
\Lc_{agg} = \frac{1}{  f_a   }\ \fr\ \ggb\ a
\ee
At low ($\Lambda_{QCD}$) energies, the $gga$ term
generates the potential $V(\tb)$ that makes 
$\tb\rightarrow 0$  and also generates the axion mass
\be
  m_a   = 
\frac{f_\pi m_\pi}{  f_a  } \frac{\sqrt{m_u m_d}}{m_u+m_d}=
0.6\ \eV\ \frac{10^7\, \GeV}{  f_a  }
\label{famarelation}
\ee
Due to these last properties the axion is not exactly a NG 
boson (which is exactly massless and has only derivative couplings). 
Notice that the axion mass is very small if the 
scale $f_a$ is very large.

The axion has also a coupling to two photons:  
\be
\Lc_{a\gamma\gamma} = c_\gamma\, \frac{\alpha}{\pi f_a }\ 
F \cdot \tilde F\ a
= -   g_{\agg}\,   \vec E \vec B\, a
\label{axionphotonphoton}
\ee
Such a coupling is important from the point of view 
of a possible detection.

Let us stress that all $c_i$ are mildly model dependent
except for the electron $c_e$ parameter. Indeed, there are 
models with $c_e=0$, i.e., the axion is not coupled 
to $e$ at tree level (KSVZ type or ``hadronic axion'') \cite{KimEM}.
However, most models have $c_e\neq0$, for example in
GUT-embedded models like the DFSZ type \cite{DineEM}.

Let us check, as a way of example, that
the $a\gamma\gamma$ coupling is not wildly model dependent.
For the DFSZ-type axion we have $c_\gamma=0.36$ and
for the KSVZ-type we have $c_\gamma=-0.97$.

\section{Limits to axion parameters}

One finds constraints on the axion properties using
laboratory experiments and astrophysical and cosmological
observations. A nice feature of the axion model is that
$f_a$ and $m_a$ are related (\ref{famarelation}), so that
there is only one   parameter in the model.
Notice that from (\ref{famarelation}) we see
that the lighter the axion is, the less interacts.

There are several high energy laboratory experiments relevant for our
discussion, like meson decays  
\ba
{\rm J}/\Psi  & \rightarrow &  \gamma a \\
\Upsilon  & \rightarrow & \gamma a \\
K^+   & \rightarrow & \pi^+ a \\
\pi^+ & \rightarrow & e^+ \nu_e a,\ \ \ a \rightarrow e^+ e^-
\ea 
beam dump experiments,  
\be
p (e^-) N \rightarrow a X\ \ \ a \rightarrow \gamma\gamma, e^+ e^-
\ee 
and nuclear deexcitation processes,  
\be
N^* \rightarrow N a \ \ \ a \rightarrow \gamma\gamma, e^+ e^- 
\ee
(whenever we consider $a \rightarrow e^+ e^- $ we obviously 
suppose $m_a>2m_e$).

The conclusion, when taking into account all these processes, is that
\be
  f_a   > 10^4\, \GeV 
\ee
or, equivalently,
\be
  m_a   < 1\, \keV
\ee
This excludes that $f_a$ could be on the order of the Fermi scale, 
which was the original suggestion of Peccei and Quinn \cite{PecceiEM}.

Astrophysical limits push very much the terrestrial limits.
The idea is that a ``too'' efficient energy drain due 
to axion emission would be inconsistent  with observation.
The most stringent limits come from horizontal branch stars in 
globular clusters. The main production is from the Primakov  
process $\gamma \gamma^* \rightarrow a$ where $\gamma^*$ corresponds
to the electromagnetic field induced by protons and electrons in
the star plasma. The coupling is restricted to \cite{RaffeltEM}
\be
  g_{\agg}   < 0.6 \times 10^{-10}\, \GeV \hspace*{.3cm} 
\Rightarrow \hspace*{.3cm}
  f_a   > 10^7\, \GeV 
\ee
In terms of axion mass, the interval
\be
0.4\, \eV\ <   m_a   < 200\, \keV
\ee
is ruled out (for $m_a >200\, \keV$ the axion is too 
heavy to be produced).

When $c_e \sim 1$ (for example, for the DSVZ axion),
the main production is from  the Compton-like process $\gamma e 
\rightarrow a e$. The stellar energy loss argument leads then
to a limit on the axion-electron coupling 
\be
  g_{aee}   \equiv c_e \frac{m_e}{  f_a  } < 2.5 \times 10^{-13} 
\ee
which enlarges the forbidden region:
\be
0.01\, \eV\ <   m_a   < 200\, \keV
\ee

The most restrictive astrophysical limits come from the analysis 
of neutrinos from
SN 1987A. In the supernova core, the main production is 
axion bremsstrahlung in nucleon-nucleon processes, 
$NN \rightarrow NN a $.

The observed duration of the $\nu$ signal at Earth detectors constrains
the coupling of the axion to nucleons. The range 
\be
3 \times 10^{-10}<
  g_{ann}   \equiv  c_n \frac{m_n}{  f_a   } 
< 3 \times 10^{-7} 
\label{limitSN}
\ee
is excluded \cite{EllisEM}. 
The upper limit in (\ref{limitSN}) corresponds to axion trapping 
in the SN. The lower limit in (\ref{limitSN}) is equivalent to
$ f_a  > 6 \times 10^{8}\ \GeV$.
In terms of the axion mass, the excluded range 
corresponding to (\ref{limitSN}) is
\be
0.01\, \eV\ <   m_a   < 10\, \eV
\ee

Other constraints from astrophysics include the ones coming
from seismic solar models \cite{WatanabeEM}, that reach the level
\be
  g_{\agg}   <  4 \times 10^{-10}\ \GeV^{-1}
\ee 
Also, white dwarfs and asymptotic giant branch stars
offer a sound astroparticle laboratories where one can get
bounds on the coupling of axions to electrons 
(see ref.\cite{DominguezEM} and the the talk of Isern 
in these Proceedings). 

Putting all the information coming from laboratory and 
astrophysics together we may conclude that the scale
of the PQ breaking is bounded by
\be
f_a  > 6 \times 10^{8}\ \GeV
\label{limitfa}
\ee

We finally consider cosmology, that puts lower limits to $m_a$. 
In the evolution of the universe, 
the cosmological history of the axion starts at temperatures
$T \sim f_a$, where $U(1)_{PQ}$ is broken. All vacuum expectation 
values $<a>$ are equally likely, but naturally we expect $<a>$
of the order of the PQ scale, or in other words an  
initial angle:   $\tb_1 \sim <a>/f_a \sim 1  $.
The next important moment in the axion history is at $T \sim$ 1 GeV,  
 since then QCD effects turn on and create a potential $V(\tb)$   
that forces $\tb \rightarrow 0$   
(CP-conserving value).

One says that the $\theta$ angle was ``misaligned'': it 
started with $  \tb=\tb_1\sim 1  $  
and will relax to   $\tb \rightarrow 0$. In the relaxation,   
the field oscillations contribute to the cosmic energy density
 \cite{PreskillEM}
\be
  \Omega h^2   
\simeq 2 \times 10^{\pm 0.4}\, F(\tb_1) \tb_1^2 \
\left( \frac{10^{-6}\, \eV}{  m_a   } \right)^{1.18}
\ee
($F$ takes into account an-harmonic effects).

This is the so-called vacuum misalignment mechanism, a process where
axions are born non-thermally and non-relativistically. 
A potentially interesting range for cosmology is 
\be
  \Omega h^2 \sim 1-0.1  \ \ \Rightarrow \ \
   m_a  \sim 10^{-3} - 10^{-6}\, \eV
\ee
since then the axion could be part of the cold dark matter 
of the universe.  

If we have as initial condition $F(\tb_1) \tb_1^2 \sim 1$, we get a 
lower bound on the axion mass
\be
10^{-6}\, \eV\, <\,    m_a  
\ee
However, for smaller values of the initial $\tb_1$, one gets a
looser bound. Apart from the value of $\tb_1$, there are 
other cosmological uncertainties.

Another axion source is the string-produced axions. 
Unless inflation occurs at $T<   f_a$,
axion strings survive and decay into axions. For many years,
there has been 
a debate on the importance of the string mechanism, and the question
is not yet settled. While a ``school'' finds 
$\Omega_{\rm string} \sim \Omega_{\rm misalign}$
another ``school'' finds
$\Omega_{\rm string} \sim 10 \, \Omega_{\rm misalign}$.
We should also mention that domain walls constitute another potential
 axion source. For a discussion of axionic strings and walls,
see ref.\cite{SikiviereviewEM}. 

For $  f_a  < 1.2 \times 10^{12}$ GeV
there is production of thermal axions in the early universe, 
but the relic density today is small
\cite{TurnerEM}
\be
n_a({\rm today}) \simeq 7.5\, {\rm cm}^{-3}  
\ee

\section{Looking for the axion}

The interaction strength of the axion scales with the inverse
of $f_a$, so that the strong bound  on $f_a$ (\ref{limitfa})
shows that the axion is a very feeble interacting particle.
The crucial observation that allows to look realistically for these 
particles was made by Sikivie \cite{SikivieEM}. The interaction
term (\ref{axionphotonphoton}) generates 
axion-photon mixing  induced by a  transverse magnetic field
$B_T$ (transverse in the sense of being perpendicular to the 
$\vec E$ polarization of the photon, the reason being the 
scalar product $\vec E \vec B$ in (\ref{axionphotonphoton}).

The mixing makes the interaction states $|a>$ and $|\gamma>$
different from the propagation states $|a'>$ and $|\gamma'>$,
\ba
|a'> &=&      \cos\varphi\, |a>\, -\, \sin\varphi\, |\gamma> \\
|\gamma'> &=& \sin\varphi\, |a>\, +\, \cos\varphi\,  |\gamma>  
\ea
The probability $P$ of the $a -\gamma$ transition is 
proportional to the small
factor $g_{\agg}^2 \sim 1/f_a^2$. However, $P$  
is enhanced when the $a -\gamma$ conversion in the magnetic
field is coherent. 
A simple way to understand coherence is to describe
the photon and the axion as plane waves propagating  
along a linear path of distance $L$. The conversion is coherent 
provided there is overlap of the wave functions across a length 
$L$, i.e.
\be
|k_{\gamma'}-k_{a'}|\, L \ll 2\pi
\label{coherent}
\ee
The probability of the coherent conversion is then
\be
  P (a \rightarrow \gamma) = \frac{1}{4}\,  
  g_{\agg}^2 \, B_T^2 \, L^2  
\ee

Searches for axions can be roughly classified in three types:
Conversion of galactic halo axions,  
conversion of solar axions
and production and detection in laboratory experiments. We
now briefly discuss each in turn. 

\subsection{Detection of halo axions}

In the presence of a galactic halo density, we expect the
conversion of axions into $\mu$-wave photons (1 GHz = 4 $\mu$eV) 
\be
h\nu=E \simeq m_a (1 + \beta^2 /2) \hspace*{1cm} \beta \sim 10^{-3}
\ee
in a cavity with a strong magnetic field (haloscope) \cite{SikivieEM}.
When the (tunable) frequency of a cavity mode equals 
the axion mass, there is a resonant conversion into radiation.
Axions are supposed to be virialized in the halo with 
$\beta \sim 10^{-3}$, so that there should be a very small dispersion.

Earlier experiments \cite{WuenschEM} put some limits, and presently 
there is already a second-generation experiment running,
the US large scale experiment\cite{AsztalosEM}, 
sensitive in the range
\be
2.9\ <   m_a   <\ 3.3\ \mu\eV
\ee
This experiment has already excluded the possibility that 
KSVZ axions 
constitute the whole of the galactic dark matter density, 
\be
\rho = 7.5 \times 10^{-25}\ {\rm g\, cm}^{-3}
\ee 
In the near future, they expect to reach $1<   m_a   < 10\, \mu\eV$
\cite{AsztalosEM}.

A promising experiment in development is CARRACK \cite{YamamotoEM}, 
in Kyoto. They will use a Rydberg atoms' technique
to detect $\mu$-wave photons.

\subsection{Detection of axions from the Sun}

A nice idea is to convert axions from the Sun 
into photons by means of a strong
magnetic field (helioscope) \cite{SikivieEM}. This search is
independent of the galactic dark matter hypothesis. The produced
photons have energies $E \sim $ a few keV (X-rays).

In Tokyo there is an experiment currently running, that has detected
no signal, giving the limit \cite{MoriyamaEM}
\be
  g_{\agg}   <  6 \times 10^{-10}\ \GeV^{-1}
\ee 
which is only valid for $m_a   < 0.03\, \eV$, to preserve coherence.

A word a caution is needed to interpret this mass $m_a$.  
In this experiment, as well as in the experiments we describe below
it is convenient to consider $g_{\agg}$ as giving
the coupling of a pseudo-scalar particle having a mass 
$m_a$, i.e., mass and coupling not related through the relation
(\ref{famarelation}). This allows to set limits with two
free parameters, $g_{\agg}$ and $m_a$. Of course, they are related
for the axion model, through the relations 
(\ref{famarelation}) and (\ref{axionphotonphoton}).

The Tokyo experiment has been improved recently by 
using gas to generate a plasmon mass
$\omega_{\rm pl}$ and thus enhancing a possible signal for 
higher particle masses, since then
$k_{\gamma'}-k_{a'} \simeq (m_a^2-\omega_{\rm pl}^2)/2E$
(see (\ref{coherent})). They get \cite{InoueEM}
\be
  g_{\agg}   <  6 \times 10^{-10}\ \GeV^{-1}
\ee
now valid for $0.05<  m_a < 0.26 \, \eV$.

In the future, a strong improvement along this line of work 
will be the CAST experiment at CERN 
 \cite{CASTEM} (see the talk of Irastorza in these Proceedings).
The experiment is ready to take data.
After two years of running, if they have found nothing, 
they expect to reach the exclusion limit
\be
  g_{\agg}   <  6 \times 10^{-11}\ \GeV^{-1}
\ee 

An alternative way to convert axions from the Sun 
into photons is to use a crystal \cite{PaschosEM}.
The needed external electromagnetic field is supplied by
the atomic Coulomb field.
There is a coherent $a \rightarrow \gamma$
conversion in a crystal
when the angle of incidence satisfies the Bragg condition. The
exclusion limits is \cite{AvignoneEM} 
\be
  g_{\agg}   < 2.7 \times 10^{-9}\ \GeV^{-1}\hspace{1cm} 
\ee
valid for $ m_a   < 1\, \keV$.

\subsection{Production and detection of axions in the laboratory}

We finally summarize the laboratory searches.
Laser light can be converted into axions when a external magnetic
field is applied and this has a variety of effects. For example,
once a photon  has converted into an axion, it can cross an opaque
substance, and after crossing it the axion may convert back into
a photon due to the magnetic field action. The net effect is 
that we have light shining through a wall. The non
observation of this phenomenon leads to the bound \cite{SemertzidisEM}
\be
  g_{\agg}  < 6.7 \times 10^{-10}\ \GeV^{-1}
\ee
for $ m_a   < 10^{-3}\, \eV $.

A very interesting idea to search for axions uses
polarized laser light \cite{MaianiEM}. 
Since the polarization $E_\parallel$ is affected by the magnetic
field but not $E_\perp$,
there are two main physical consequences. The axion can be
produced so that there is
a selective absorption of $E_\parallel$. This is called
dichroism and produces a rotation of the polarization plane.
The second effect is birefringence, since when there is virtual 
production of an axion,
the index of refraction for $E_\parallel$ is different
than for $E_\perp$.

The PVLAS experiment \cite{BrandiEM} constructed to search for these
effects is now running. They get some signal and are analyzing whether
it comes from some new physics. In any case, they expect to be sensitive
to
\be  
g_{\agg}  \sim  10^{-7}\, \GeV^{-1}
\label{pvlas}
\ee
for
\be 
  m_a   < 10^{-3}\ \eV
\ee
Although the figure (\ref{pvlas})
is above other limits we have been quoting, we should
stress that this experiment is independent of 
the galactic dark matter and solar production hypothesis.

The PVLAS experiment \cite{BrandiEM} is also interesting 
for ``conventional physics'',
because when improving the sensitivity, they should ``see''
the birefringence in vacuum generated by the  QED 
light-light box-diagram.

\section{Axion-like particles}

Global symmetries that are spontaneously broken lead
to NG bosons. An example is  
family symmetry, which would be related to the
number and properties of families (we still don't have an answer to
Rabi's question:  Who ordered the muon?). The breaking of the symmetry
would  give rise to familons. Another example is    
lepton-number symmetry, that would produce majorons. There are more
theoretical examples, so the axion is not the only NG
boson that has been proposed. In general, all these bosons couple
to photons, so that they could give signatures in most of the experiments
that look for axions.

It is then interesting to analyze experimental constraints
with two free parameters: mass and coupling to $\gamma\gamma$.
 To do that, we make the hypothesis that
there is a boson $\varphi$ with mass $m$ and coupling $g$
\be
\Lc =  \frac{1}{8}\ g\  \epsilon_{\mu \nu \alpha \beta} 
F^{\mu \nu} F^{\alpha \beta}\ \varphi
\ee
Experimental constraints for $g$ as a function of $m$ 
are presented in \cite{MassoEM}.

Another reason to relax the relation (\ref{famarelation}) is that
it could be no longer valid in axion model where there are
contributions to axion mass from exotic sources.

There are still many open questions in the field of
hypothetical NG bosons. For example, the
coupling of familons to the third family is
very poorly constrained \cite{FengEM}. Also, there may be
unexpected   possibilities, like for example the suggestion
of Ref.\cite{CsakiEM}, where they show that an
axion-like $\varphi$ particle with
\be
m \sim 10^{-16}\ {\rm eV}
\ee
and
\be
g \sim 2 \times 10^{-12}\ {\rm GeV}^{-1}
\ee
would make that 20-30\% of photons from distant SNe
oscillate into axion-like particles in presence of an extra-galactic 
magnetic field $B\sim 10^{-9}$ G, contributing 
to the dimming of SNe. Although it is highly speculative,
the consequences for the measurements of the acceleration of
the universe are very exciting. As a final example, let us mention
that in Ref.\cite{GrifolsEM} the production of pseudoscalars
in very strong electromagnetic fields has been analyzed.

\section{Conclusions}

The axion was born as a consequence of the PQ
elegant solution to the strong CP problem. It has 
quite precise properties, with just one free
parameter. Laboratory, astrophysical and cosmological
observations constrain the axion parameter. The upcoming
experiments may find the axion, or may exclude it.
Descendants of axions, that we call
axion-like bosons, also offer a piece of interesting physics.

\begin{acknowledgments}
Work partially supported by the CICYT Research Project AEN99-0766, 
by the EU network on Supersymmetry and the Early Universe 
(HPRN-CT-2000-00152), and by the
{\it Departament d'Universitats, Recerca i Societat de la Informaci{\'o}},
Project 2001SGR00188.
\end{acknowledgments}


\begin{thebibliography}{9}
\bibitem{PecceiEM}
R.~D.~Peccei and H.~R.~Quinn,
Phys.\ Rev.\ Lett.\  {\bf 38} (1977) 1440.
\\
R.~D.~Peccei and H.~R.~Quinn,
Phys.\ Rev.\ D {\bf 16} (1977) 1791.


\bibitem{WeinbergEM}
S.~Weinberg,
Phys.\ Rev.\ Lett.\  {\bf 40} (1978) 223.
\\
F.~Wilczek,
Phys.\ Rev.\ Lett.\  {\bf 40} (1978) 279.

\bibitem{ChengEM}
For a review, see H.~Y.~Cheng,
Phys.\ Rept.\  {\bf 158} (1988) 1.

\bibitem{KimEM}
J.~E.~Kim,
Phys.\ Rev.\ Lett.\  {\bf 43} (1979) 103.
\\
M.~A.~Shifman, A.~I.~Vainshtein and V.~I.~Zakharov,
Nucl.\ Phys.\ B {\bf 166} (1980) 493.

\bibitem{DineEM}
M.~Dine, W.~Fischler and M.~Srednicki,
Phys.\ Lett.\ B {\bf 104} (1981) 199.
\\
A.~R.~Zhitnitsky,
Sov.\ J.\ Nucl.\ Phys.\  {\bf 31} (1980) 260
[Yad.\ Fiz.\  {\bf 31} (1980) 497].

\bibitem{RaffeltEM}
G.~G.~Raffelt,
``Stars As Laboratories For Fundamental Physics: 
The Astrophysics Of Neutrinos, Axions, And Other 
Weakly Interacting Particles,''
{\it  Chicago, USA: Univ. Pr. (1996) 664 p}.

\bibitem{EllisEM}
J.~R.~Ellis and K.~A.~Olive,
Phys.\ Lett.\ B {\bf 193} (1987) 525.\\
G.~Raffelt and D.~Seckel,
Phys.\ Rev.\ Lett.\  {\bf 60} (1988) 1793.\\
M.~S.~Turner,
Phys.\ Rev.\ Lett.\  {\bf 60} (1988) 1797.

\bibitem{WatanabeEM}
S.~Watanabe and H.~Shibahashi,
arXiv:hep-ph/0112012.

\bibitem{DominguezEM}
I.~Dominguez, O.~Straniero, and J.~Isern,
MNRAS, {\bf 306} (1999) L1.


\bibitem{PreskillEM}
J.~Preskill, M.~B.~Wise and F.~Wilczek,
Phys.\ Lett.\ B {\bf 120} (1983) 127.
\\
L.~F.~Abbott and P.~Sikivie,
{\it ibid.} 133.
\\
M.~Dine and W.~Fischler,
{\it ibid.} 137.
\\
M.~S.~Turner,
Phys.\ Rev.\ D {\bf 33} (1986) 889.

\bibitem{SikiviereviewEM}
P.~Sikivie,
Nucl.\ Phys.\ Proc.\ Suppl.\  {\bf 87} (2000) 41.

\bibitem{TurnerEM}
M.~S.~Turner,
Phys.\ Rev.\ Lett.\  {\bf 59} (1987) 2489
[Erratum-ibid.\  {\bf 60} (1987) 1101].
\\
E.~Masso, F.~Rota and G.~Zsembinszki,
Phys.\ Rev.\ D {\bf 66} (2002) 023004.


\bibitem{SikivieEM}
P.~Sikivie,
Phys.\ Rev.\ Lett.\  {\bf 51} (1983) 1415
[Erratum-ibid.\  {\bf 52} (1984) 695].

\bibitem{WuenschEM}
W.~U.~Wuensch {\it et al.},
Phys.\ Rev.\ D {\bf 40} (1989) 3153.
\\
C.~Hagmann, P.~Sikivie, N.~S.~Sullivan and D.~B.~Tanner,
Phys.\ Rev.\ D {\bf 42} (1990) 1297.

\bibitem{AsztalosEM}
S.~Asztalos {\it et al.},
Phys.\ Rev.\ D {\bf 64} (2001) 092003.

\bibitem{YamamotoEM}
K.~Yamamoto {\it et al.},
arXiv:hep-ph/0101200.

\bibitem{MoriyamaEM}
S.~Moriyama, M.~Minowa, T.~Namba, Y.~Inoue, Y.~Takasu and A.~Yamamoto,
Phys.\ Lett.\ B {\bf 434} (1998) 147.


\bibitem{InoueEM}
Y.~Inoue, T.~Namba, S.~Moriyama, M.~Minowa, Y.~Takasu, 
T.~Horiuchi and A.~Yamamoto,
Phys.\ Lett.\ B {\bf 536} (2002) 18.


\bibitem{CASTEM}
Home page:\\ 
http://axnd02.cern.ch/CAST/

\bibitem{PaschosEM}
E.~A.~Paschos and K.~Zioutas,
Phys.\ Lett.\ B {\bf 323} (1994) 367.

\bibitem{AvignoneEM}
F.~T.~Avignone {\it et al.}  [SOLAX Collaboration],
Phys.\ Rev.\ Lett.\  {\bf 81} (1998) 5068.
\\
A.~Morales {\it et al.}  [COSME Collaboration],
Astropart.\ Phys.\  {\bf 16} (2002) 325.

\bibitem{SemertzidisEM}
Y.~Semertzidis {\it et al.},
Phys.\ Rev.\ Lett.\  {\bf 64} (1990) 2988.

\bibitem{MaianiEM}
L.~Maiani, R.~Petronzio and E.~Zavattini,
Phys.\ Lett.\ B {\bf 175} (1986) 359.

\bibitem{BrandiEM}
F.~Brandi {\it et al.},
Nucl.\ Instrum.\ Meth.\ A {\bf 461} (2001) 329.
\\
G. Cantatore, talk at
4th International Workshop On The Identification Of Dark Matter 
(IDM2002), York, England.

\bibitem{MassoEM}
E.~Masso and R.~Toldra,
Phys.\ Rev.\ D {\bf 52} (1995) 1755.
\\
E.~Masso and R.~Toldra,
Phys.\ Rev.\ D {\bf 55} (1997) 7967.
\\
J.~A.~Grifols, E.~Masso and R.~Toldra,
Phys.\ Rev.\ Lett.\  {\bf 77} (1996) 2372.
\\
J.~W.~Brockway, E.~D.~Carlson and G.~G.~Raffelt,
Phys.\ Lett.\ B {\bf 383} (1996) 439.

\bibitem{FengEM}
J.~L.~Feng, T.~Moroi, H.~Murayama and E.~Schnapka,
Phys.\ Rev.\ D {\bf 57} (1998) 5875.

\bibitem{CsakiEM}
C.~Csaki, N.~Kaloper and J.~Terning,
Phys.\ Rev.\ Lett.\  {\bf 88} (2002) 161302.

\bibitem{GrifolsEM}
J.~A.~Grifols, E.~Masso, S.~Mohanty and K.~V.~Shajesh,
Phys.\ Rev.\ D {\bf 65}, 055004 (2002).\\
J.~A.~Grifols, E.~Masso, and S.~Mohanty,
Phys.\ Rev.\ D {\bf 60}, 097701 (1999)
[Erratum-ibid.\ D {\bf 65}, 099905 (2002)].

\end{thebibliography}
\end{document}